\documentclass[dvips]{article}

\usepackage{icrc2011}
\usepackage{stfloats}

\newcommand{\g}{$\gamma$}
\newcommand{\F}{\textit{Fermi}}
\newcommand{\xco}{X_\mathrm{CO}}
\newcommand{\hi}{\mathrm{H\,\scriptstyle{I}}}
\newcommand{\nhi}{N(\mathrm{H\,\scriptstyle{I}})}

\newcommand{\wco}{W_\mathrm{CO}}
\newcommand{\nhd}{N(\mathrm{H_2})}

\newcommand{\av}{\mathrm{A}_\mathrm{V}}

\newcommand{\adeg}{^\circ}
\newcommand{\degr}{\adeg}

\newcommand{\qhi}{q_\mathrm{H\,\scriptscriptstyle{I}}}
\newcommand{\qco}{q_\mathrm{CO}}

\newcommand{\qdust}{q_\mathrm{\mathrm{A}_\mathrm{V}}}

\newcommand{\xav}{X_\mathrm{A_V}}
\newcommand{\avres}{\mathrm{A}_\mathrm{V,exc}}

\title{The \F-LAT view of cosmic rays and interstellar gas in the Cygnus region: a not so
special spot of the Local Arm}

\newcommand{\etal}{\MakeLowercase{\textit{et al. }}} 
\shorttitle{Tibaldo~L. \etal Interstellar gamma-ray emission from Cygnus}

\authors{Luigi Tibaldo$^{1}$, Isabelle~A. Grenier$^{2}$ on behalf of the \F{} LAT
collaboration}
\afiliations{$^1$Dipartimento di Fisica ``G. Galilei'', Universit\`a di Padova, and INFN-Sezione
di Padova, Italy\\  $^2$AIM--Universit\'e Paris Diderot and CEA Saclay, France}
\email{luigi.tibaldo@pd.infn.it}

\abstract{The Cygnus region hosts the most conspicuous star-forming region close to the
Sun, embedded in a
giant complex of molecular clouds in the Local Arm. We present an analysis of the
\F-LAT observations of Cygnus intended to probe the cosmic-ray and
interstellar-matter content of the region. From gamma-ray data we estimate a total of
$8^{+5}_{-1} \times 10^6 M_\odot$ of interstellar gas in the complex at a
distance of 1.4
kpc. The gamma-ray emission from the atomic gas supports the average $\hi$ spin
temperature derived from radio absorption/emission pairs to estimate its
column densities. The $\xco=\nhd/\wco$ ratio derived in the massive Cygnus complex is
consistent with other LAT estimates for clouds in the Local and Perseus arms. The
mass of dark gas, escaping $\hi$ and CO observations but traced by dust and gamma rays,
amounts to $\sim 10\%$ of the total. We find an average gamma-ray
emissivity per interstellar H atom in the 0.1--100 GeV energy band in good agreement
with measurements in other segments of the Local Arm. We infer that the cosmic-ray
population averaged over a few hundred parsecs is fairly uniform in density and
spectrum along the Local Arm. Despite the presence of potential accelerators and much
larger interstellar densities in Cygnus compared to the solar neighborhood, their
cosmic-ray populations are similar on such a scale.}

\begin{document}
\maketitle

\section{Introduction}
Interstellar \g-ray emission produced by cosmic-ray (CR) interactions with interstellar
gas, through nucleon-nucleon collisions and electron Bremmstrahlung, is a tracer of the product of
CR and interstellar-matter densities throughout the Galaxy. The data which are being collected by
the \F{} Large Area Telescope (LAT) \cite{ref:LAT} are therefore providing a wealth of new
information on the
interstellar environment of the Milky Way.

We used LAT data to characterize the CR and gas content of the Cygnus region, which is located
around the Galactic longitude $l=80\degr$, tangent to the Local Arm. A super-massive molecular
complex harbors the region of massive star formation known as Cygnus~X, at $\sim 1.4$~kpc from the
solar system \cite{ref:negueruela2008}. Beyond the Local Arm two segments of the Perseus and outer
spiral arms are seen in this direction.

In this contribution we present the global model that was sought for the high-energy \g-ray emission
from Cygnus
and we discuss the
implications for the properties of CRs and the census of the interstellar medium (ISM) over the
scale of the whole complex, i.e. a few hundred parsecs. The innermost Cygnus~X region will be
discussed in another contribution to this conference \cite{ref:cocoon}.

\section{LAT Data Analysis}

\subsection{\g-ray Data}\label{data}
We used the first two years of the LAT survey applying stringent event selection criteria to limit
the background due to misclassified CR interactions in the LAT \cite{ref:IGB} and the contamination
from the Earth's atmospheric emission. We analysed LAT data in the region at Galactic longitude
$72\degr \leq l \leq 88\degr$ and latitude $|b| \leq 15\degr$. Photons were used to construct
maps on a $1/8\degr$ angular grid and over ten bins in energy from 100~MeV to 100~GeV. 

The \g-ray emission from this region is dominated at energies $< 10$~GeV by three bright
pulsars: J2021+3651, J2021+4026 and J2032+4127. To increase the sensitivity to diffuse and fainter
extended sources at these energies for each pulsar we removed from our data sample the photons
detected in the periodic time intervals where the pulsed emission peaks, in a region around the
source position with a radius approximately equivalent to the acceptance-averaged $95\%$
point-spread function (PSF) of the LAT as a function of energy.

\subsection{Analysis Model}\label{model}
Since the Galaxy is transparent to high-energy \g-rays and the CR densities are thought to be
uniform over the characteristic dimensions of interstellar complexes, we can model to first
order emission from gas as a linear combination of gas column densities for the different complexes
along the line of sight and the different phases of the ISM.

Column densities of atomic hydrogen, $\hi$, were derived from the observations of its 21-cm line
\cite{ref:CGPS,ref:LAB}. Column densities of molecular gas are assumed to be proportional to the
integrated intensities of the 2.6-mm line, $\wco$ \cite{ref:CO,ref:COmom}, $\nhd=\xco \cdot \wco$.
$\hi$ and CO data were
used to build maps of gas column densities partitioned along the line of sight to separate the
Cygnus complex from the spiral arms in the outer Galaxy, according to the method described in
\cite{ref:IIquad,ref:IIIquad}.

The dark neutral gas which is missed by $\hi$ and CO observations was recovered by a dust excess
map \cite{ref:darkgas}, namely by the visual extinction $\av$ estimated from the 2MASS source
Catalog \cite{ref:av1,ref:av2,ref:av3} minus the best-fit linear combination of the afore described
$\hi$ and CO maps.

The emission from interstellar gas is therefore modeled as a linear combination of the $\hi$, CO
and $\avres$ maps with free combination coefficients independently in each energy bin.
These combination coefficients are the emissivities, i.e. \g-ray emission rates, per
H atom, $\qhi$, per $\wco$ unit, $\qco$, and per $\avres$ unit, $\qdust$.

The interstellar emission model is completed by the large-scale Galactic inverse-Compton emission
produced by CR electrons interacting with low-energy radiation fields, which was modeled using the
GALPROP code\footnote{\texttt{http://galprop.stanford.edu/}}, run \texttt{54\_87Xexph7S}. An
isotropic background component, with free normalization in each independent energy bin, was added
to take into account the isotropic \g-ray emission and the residual background from misclassified
CR interactions in the LAT. 

Several sources are included in the model: all the sources in the first LAT source
Catalog (1FGL) which are identified, associated at high confidence level with Active Galactic
Nuclei or show strong variability. We then introduced some extended sources detected with LAT data:
the supernova remnants known as the Cygnus Loop \cite{ref:CygLoop} and \g~Cygni, the source
VER~J2019+407 \cite{ref:veritas} and the extended source in the innermost part of the Cygnus~X
region discussed in \cite{ref:cocoon}. The remainder of the three bright pulsars, phase-suppressed
as
described in \S~\ref{data}, was modeled for each of them by a point source to take into account the
off-pulse
emission and by a second source, for which, after folding with the instrument response as described
in \S~\ref{ana}, expected counts were set to null within the exclusion
radius introduced in \S~\ref{data}, to take into account the spill-over in the PSF tails. For each
source the flux was let as a free parameter independently in each energy bin.

\subsection{Analysis Method}\label{ana}
The standard LAT analysis
software\footnote{\texttt{http://fermi.gsfc.nasa.gov/ssc/data/\\analysis/documentation/Cicerone/}}
was used to calculate the energy-dependent exposure, which was then scaled to take into account the
selection on pulsar phases (\S~\ref{data}). The analysis model described above was corrected for
the exposure and convolved with the PSF, to obtain the number of counts expected from the model in
each pixel of the maps. This was then fit to the LAT data by a maximum-likelihood method based on
Poisson statistics in order
to determine the best model parameters.

The analysis was performed using the Instrument Response Functions (IRFs) of the series
\texttt{P6\_V3} \cite{ref:rando}. The relative systematic uncertainties described therein are
taken into account in the interpretation of the results.

\section{Results and Discussion}

\subsection{The Issue of the $\hi$ Spin Temperature}\label{TS}
\begin{figure}[!htbp]
  \vspace{5mm}
  \centering
  \includegraphics[width=0.5\textwidth]{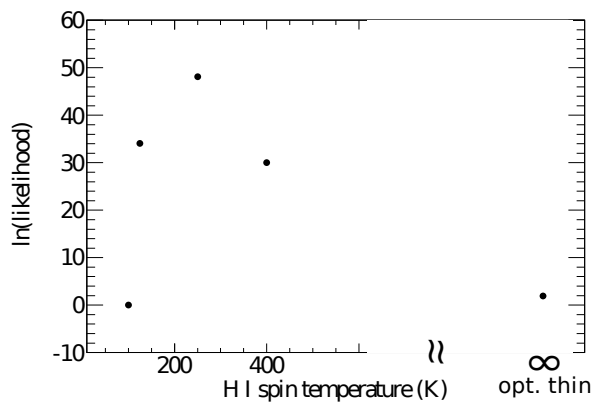}
  \caption{PRELIMINARY -- Maximum likelihood obtained by fitting the model to the LAT data for
some discrete values of the $\hi$ spin temperature $T_S$ used to prepare $\nhi$ maps: 100~K, 125~K,
250~K, 400~K and the limit of infinitely high $T_S$ (optically thin medium).}
  \label{fig:Ts}
 \end{figure}

The preparation of the $\nhi$ maps described in \S~\ref{model} requires to account for the optical
depth of the gas, which is usually done by assuming a uniform value of the so-called $\hi$ spin
temperature,
$T_S$. No simple solutions of the $\hi$-line transfert equation exist beyond the approximation of
uniform $T_S$. We therefore explored a wide range of possible average temperatures from 100~K to
$\infty$ (optically
thin medium). The corresponding maximum-likelihood values obtained from the \g-ray analysis are
shown in Fig.~\ref{fig:Ts}.

The results support the average $T_S$ values of a few hundred K estimated from radio
absorption-emission pairs \cite{ref:dickey} to estimate the $\hi$ column densities. The spin
temperature, however, is known to be not
uniform: we used the different values explored to gauge the subsequent uncertainties affecting
the results.

\subsection{CR Densities in Cygnus and along the Local Arm}

\begin{figure}[!htbp]
  \vspace{5mm}
  \centering
  \includegraphics[width=0.5\textwidth]{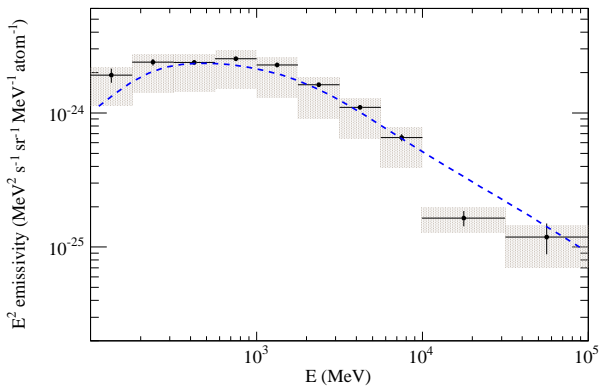}
  \caption{PRELIMINARY -- Spectrum of emissivity per H atom in Cygnus. Shaded
bands
indicate the systematic uncertainties, combining the imperfect knowledge of the instrument response
and of the $\hi$ optical depth. The dashed line shows the expectations based on a CR
spectrum
consistent with direct measurements near the Earth \cite{ref:hiemiss}.}
  \label{fig:hispec}
 \end{figure}

Fig.~\ref{fig:hispec} shows the spectrum of \g-ray emissivity per H atom in the Cygnus complex,
which is consistent with the predictions based on the CR spectra directly measured near the Earth
and
\g-ray emission from the nearby interstellar space as observed by the LAT \cite{ref:hiemiss}.

\begin{figure}[!t]
  \vspace{5mm}
  \centering
  \includegraphics[width=0.5\textwidth]{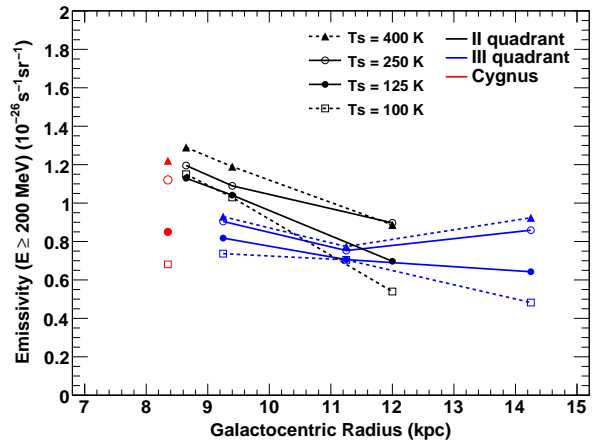}
  \caption{PRELIMINARY -- Emissivity per H atom integrated for $E>200$~MeV as a function of
Galactocentric radius, combining results from this analysis (red, or light gray) and from the
studies of the second \cite[black]{ref:IIquad} and third \cite[blue or dark gray]{ref:IIIquad}
Galactic quadrants. Statistical errors are comparable with marker dimensions, whereas the different
markers show the results for different values of
the $\hi$ spin temperature $T_S$ used to prepare $\nhi$ maps.}
  \label{fig:emprof}
 \end{figure}

Fig.~\ref{fig:emprof} compares the $\hi$ emissivity integrated above 200~MeV in the Cygnus complex
and other segments of the Local Arm and regions toward the outer Galaxy
\cite{ref:IIquad,ref:IIIquad,ref:outgal}.
Given the large uncertainties due to the issue of the $\hi$
spin temperature (\S~\ref{TS}), Fig.~\ref{fig:emprof} shows that the CR population averaged over a
scale of a few hundred pc is uniform within $10\%-35\%$ along the Local Arm.

\begin{figure*}[!btp]
   \centerline{\includegraphics[width=0.5\textwidth]{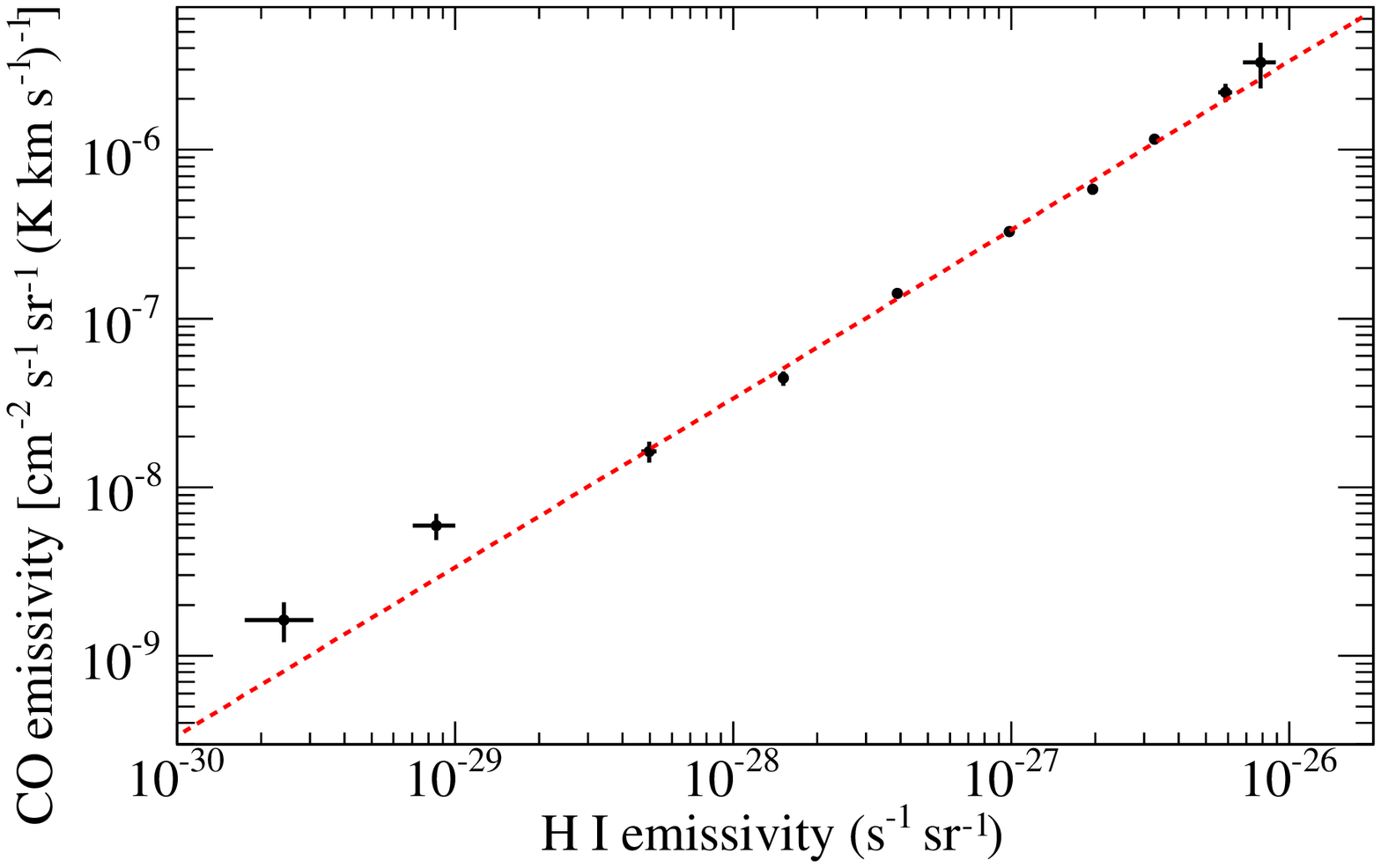}
              \hfil
              \includegraphics[width=0.5\textwidth]{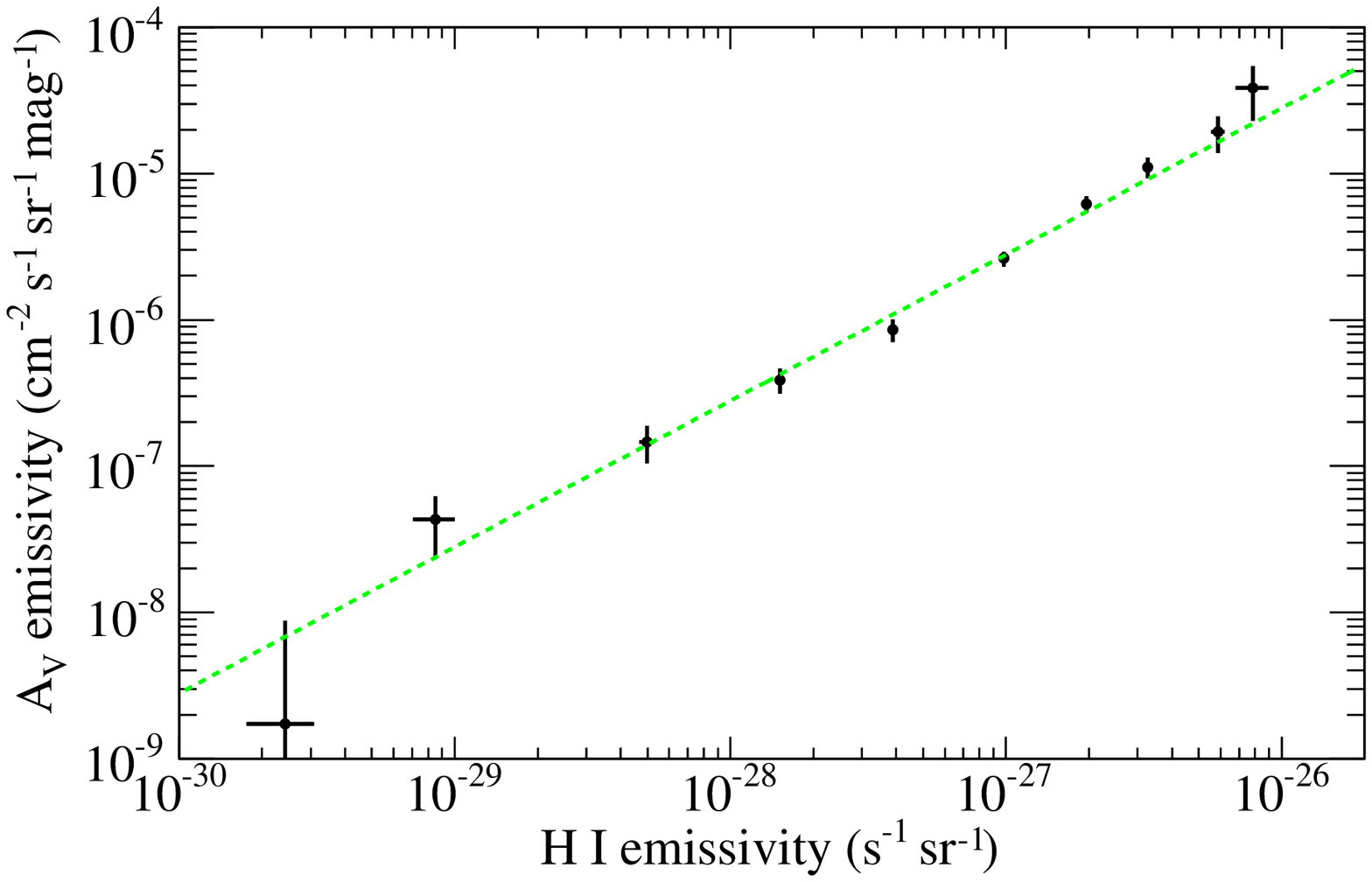}
             }
   \caption{PRELIMINARY -- Emissivity per $\wco$ unit (left) and per $\avres$ unit (right) versus
emissivity per H atom for the different energy bins considered in the analysis. The dashed lines
represent the best-fit linear
relations taking into account uncertainties on both axes.}
   \label{fig:emisscat}
 \end{figure*}

\subsection{The Census of the Interstellar Medium}   

By the definition $\nhd=\xco \cdot \wco$, we obtain $q_\mathrm{H_2}=\xco^{-1}
\cdot \qco$. Assuming that the CR densities are the same in all the
phases of an interstellar complex, we expect $q_\mathrm{H_2}=2\qhi$. We
therefore expect the emissivity per $\wco$ unit to linearly scale as
the emissivity per H atom, $\qco=2\xco \cdot \qhi$. This is verified in
Fig.~\ref{fig:emisscat}~(left), with good confidence
in the 0.1--10 GeV energy range. A small ($<3\sigma$) deviation is seen $>10$~GeV (i.e. at low
emissivities), possibly due to the hard extended emission \cite{ref:cocoon} observed in the
innermost Cygnus~X region where most of the CO-bright gas is located. From the slope of the
best-fit linear relation we can derive $\xco$ and estimate the masses of CO-bright gas.

We similarly expect the emissivity per $\avres$ unit to linearly scale as
the emissivity per H atom, as we verified in Fig.~\ref{fig:emisscat}~(right). The linear
correlation implies that \g-ray emission associated to dust excesses and atomic gas is produced by
the same physical processes, i.e. that $\avres$ indeed trace interstellar gas. From the best-fit
linear relation we can derive the dust-to-gas ratio in the dark neutral phase,
$\xav=N(\mathrm{H})/\avres$, and estimate its mass. Whether the dark
neutral gas is CO-quiet molecular gas or atomic gas missed because of the approximations in the
handling of the radiative transfert of the $\hi$ line should be investigated in the future.

Assuming a distance of 1.4 kpc from the solar system, the total masses estimated on the basis on the
LAT data analysis amount to $8^{+5}_{-1} \times 10^6
M_\odot$ in the Cygnus complex, of which
$\sim 60\%$ in the form of atomic gas, $\sim 30\%$ in the form of CO-bright molecular gas and $\sim
10\%$ in the form of dark neutral gas.

\begin{figure}[!tb]
  \vspace{5mm}
  \centering
  \includegraphics[width=0.5\textwidth]{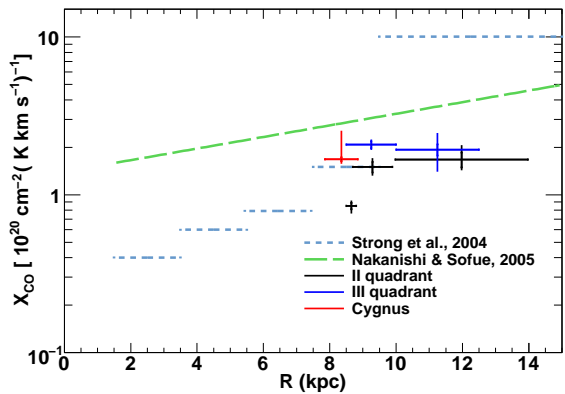}
  \caption{PRELIMINARY -- $\xco$ as a function of Galactocentric radius. Points represent LAT
measurements for the Cygnus clouds (red or light gray) and several complexes seen in the second
\cite[black]{ref:IIquad} and third \cite[blue or dark gray]{ref:IIIquad} Galactic quadrants. Some
models reported in the literature \cite{ref:nakanishi2006,ref:strong2004} are shown for
comparison.}
  \label{fig:xco}
 \end{figure}

The $\xco$ ratio obtained for the Cygnus complex appears consistent with estimates from LAT
data in other segments of the Local Arm and in the Galactic disc over a few kpc
beyond the solar circle (Fig~\ref{fig:xco}). 
It is significantly larger than the measurement for the nearby clouds of Cassiopeia and Cepheus
\cite{ref:IIquad};
the origin of this difference will be investigated in the future.

\section{Summary}

The Cygnus region appears as a very peculiar segment of the Local Arm because of the large masses
of interstellar gas, confirmed by its \g-ray emission, and the intense massive-star
formation activity, and therefore the potential
presence of numerous particle accelerators. Despite these peculiarities, the properties of CRs over
the scale of the whole complex are similar to
those of other segments of the Local Arm studied with LAT data.


\vspace{\baselineskip}
\subsubsection*{Acknowledgments}
The \F{} LAT Collaboration acknowledges support from a number of agencies and institutes for
both development and the operation of the LAT as well as scientific data analysis. These include
NASA and DOE in the United States, CEA/Irfu and IN2P3/CNRS in France, ASI and INFN in Italy, MEXT,
KEK, and JAXA in Japan, and the K.~A.~Wallenberg Foundation, the Swedish Research Council and the
National Space Board in Sweden. Additional support from INAF in Italy and CNES in France for science
analysis during the operations phase is also gratefully acknowledged.

\clearpage

\end{document}